**Histopathology of Third Trimester Placenta from SARS-CoV-2-Positive Women**

Mai He, M.D., Ph.D.[1]*, Priya Skaria, M.D. [1], Kasey Kreutz, M.D. [1], Ling Chen, Ph.D.[2], Ian Hagemann, M.D., Ph.D. [1], Ebony B. Carter, M.D.[3], Indira U. Mysorekar, Ph.D. [1,3], D Michael Nelson, M.D., PhD. [3], John Pfeifer, M.D., Ph.D. [1], Louis P. Dehner, M.D.[1]

1. Department of Pathology & Immunology, Washington University in St. Louis School of Medicine, St. Louis, MO 63110, USA
2. Division of Statistics, Washington University in St. Louis School of Medicine, St. Louis, MO 63110, USA
3. Department of Obstetrics & Gynecology, Washington University School of Medicine in St. Louis, St. Louis, MO 63110, USA

Corresponding author*

Mai He, M.D., Ph.D.

Associate Professor

Department of Pathology & Immunology,

Washington University in St. Louis School of Medicine,

St. Louis, MO 63110, USA

Phone: (314) 273-1328

Email: Maihe@wustl.edu

Orcid: 0000-0002-4775-9757

**Abstract 145**

**Background:** This study aims to investigate whether maternal SARS-CoV-2 status affect placental pathology. **Methods:** A retrospective case-control study was conducted by reviewing charts and slides of placentas between April 1 to July 24, 2020. Clinical history of "COVID-19" were searched in Pathology Database (CoPath). Controls were matched with SARS-CoV-2-negative women with singleton deliveries in the 3$^{rd}$-trimester. Individual and group, pathological features were extracted from placental pathology reports. **Results:** Twenty-one 3$^{rd}$-trimester, placentas from SARS-CoV-2-positive women were identified and compared to 20 placentas from SARS-CoV-2-negative women. There were no significant differences in individual or group gross or microscopic pathological features between the groups. Within the SARS-CoV-2+ group, there are no differences between symptomatic and asymptomatic women. **Conclusion:** Placentas from SARS-CoV-2-positive women do not demonstrate a specific pathological pattern. Pregnancy complicated with COVID-19 during the 3$^{rd}$ trimester does not have a demonstrable effect on placental structure and pathology.

## Introduction

Coronavirus disease 2019 (COVID-19) is caused by infection of SARS-CoV-2, a member of the betacoronavirus family. SARS-CoV-2 is a 30 kb enveloped, positive sense, single-stranded RNA virus. The virus consists of four structural proteins (spike surface glycoprotein, envelope protein, membrane protein, and nucleocapsid protein) and non-structural proteins. The spike protein consists of two functional subunits. The S1 subunit is responsible for binding to the host cell receptor and the S2 subunit is utilized for the fusion of the viral and cellular membranes.[1,2] Studies showed that the spike protein for SARS-CoV-2 binds to angiotensin converting enzyme 2 (ACE2), which is also a functional receptor for SARS-CoV. ACE2 expression is high in lung, heart, ileum, kidney and bladder. Thus, SARS-CoV-2 virus primarily affects the respiratory system, although other organ systems are also involved.[1]

Approximately 1% of infected people develop severe acute respiratory distress syndrome (ARDS) that requires a critical level of care. A MMWR study suggested that pregnant women with COVID-19 are more likely to be hospitalized, and are at increased risk for intensive care unit (ICU) admission and receipt of mechanical ventilation, compared to nonpregnant women.[3] While much remains unknown, affected respiratory function affects raise concerns regarding oxygenation during pregnancy.

Studies of pregnancy outcomes associated with COVID-19 infection suggest that pregnant women with SARS-CoV2 infection have increased pregnancy complications.[4,5] The placenta is the key organ at the maternal/fetal interface and is known to be the root cause for some pregnancy complications. The placenta essentially functions as the fetal lungs and kidneys and has been implicated in SARS-CoV-2 infection. Valdés G et al have reported immunocytochemical expression of ACE2 in the syncytiotrophoblast (ST), cytotrophoblast (CT), endothelium and vascular smooth muscle of primary and secondary villi, [6] while Vivanti A et al demonstrated potential evidence of transplacental transmission.[7] Thus, it is important to study the effects of SARS-CoV2 infection on placental structure and growth.

We tested the null hypothesis that placentas in women infected at the mid-to-late stage of pregnancy would show no impact of the viral infection on placental structure and growth.

## Materials and Methods

*Institution Review Board (IRB) approval*

IRB ID# 201902092 was approved by the Office of Institution Review Board, Washington University in St. Louis.

*Study design*

Retrospective case-control review charts and slides during April 1 to July 24, 2020.

*Study Patients*

Placental reports from pregnant women with positive SARS-CoV-2 testing were identified via the electronic health record in CoPath. They were all third trimester singleton deliveries during April 1st to July 24, 2020.

*Control group*

Control placentas were chosen from deliveries during April 1st to July 24, 2020. The pregnant women chosen were tested for SARS-CoV-2 with negative results. The first third trimester singleton placenta, accessioned in Histology Laboratory (in the Pathology department) after the identified SARS-CoV-2 + placenta, were used as study controls. The exclusion criteria were those without testing, with fetal anomalies, or with a multiple gestation pregnancy.

*Clinical information*

*C*linical information was extracted from the electronic medical records (EPIC).

*Placental pathology review and central review*

All placentas were grossly examined and microscopically examined with H&E-stained sections, which usually included one section with two umbilical cord sections and membrane rolls, and three full-thickness sections that included fetal and maternal surfaces. Additional sections were submitted for any tissue with presumed lesions. The placental pathology reports were generated by Anatomic or Pediatric Pathology (PediP) board certified pathologists following established criteria and were centrally reviewed by PediP board-certified pathologist.[8,9] The slides for every case were reviewed by two pathologists.

*SARS-CoV-2 testing*

Testing for pregnant women via nasopharyngeal swabs was performed at Barnes Jewish Hospital Molecular Infectious Disease Laboratory. The 2019-Novel Coronavirus Assay (COVID-19) Real Time RT-PCR assay was used to detect the presence of SARS-CoV-2 RNA.

*Statistics*

The Shapiro–Wilks test was used to check normality of the distribution of the placental morphometric variables. The mean and standard deviation of each placental morphometric variable were obtained and mean comparisons between groups were performed using a 2-sample t-test or Mann–Whitney U test, as appropriate. Chi-square or Fisher's exact test was used as appropriate for categorical variables. Significance was set at $p<0.05$. All the statistical tests were two-sided and performed with SAS 9.4 (SAS Inc, Cary, NC).

**Results**

A computerized Copath search yielded 21 placentas from pregnant women who tested positive for SARS-CoV-2 (the study group), all delivered in the third trimester between 33-40 weeks. Correspondingly, 20 control third trimester placentas were chosen by criteria mentioned in Methods (the control group). Clinical information is summarized in Table 1. There were no significant differences in maternal age, gestational age at delivery, placental weight, or fetal/placental (F/P) weight ratio between the two groups.

Detailed histopathological features of the study group are summarized in Table 2. Among the 21 patients tested positive for COVID-19, seven (7/21, 33.3%) had respiratory symptoms. Patient #15 delivered at 34 weeks' gestation and the infant died in seven hours. Fourteen (14/21, 66.7%) placentas had weights that were < $10^{th}$ % for gestational age, and thus, were small for gestational age There were no large for gestational age placentas.

Placentas from SARS-CoV-2-positive women did not demonstrate specific pathology or pathological pattern. There were no significant differences in individual or group histopathological features between the study and control groups. In the study group, ten of 21 placentas (47.6%) had features of maternal vascular malperfusion (MVM) which was not significantly different from the four of 20 placentas (20%) with MVM in the control group. Between the study and control groups, there were ten of 21 (47.6%) placentas in the study group showing any features of fetal vascular malperfusion (FVM), not significantly different from the eight of 20 control placentas (40%) with FVM features. There were two of 21 study group, and three of 20 control, placentas showing maternal inflammatory response, and two of 21 study group, and four of 20control, placentas showing the fetal inflammatory response (). Each of the two groups had one placenta with villitis of unknown etiology (VUE) and four placentas with chronic deciduitis with plasma cells. A single placenta, in the study group, exhibited T/eosinophilic vasculitis. None of these pathologies were significantly different between the two groups. There were no differences in abnormal cord insertion or cord coiling between the two groups, not any differences in chorioamnion pathology.

Within the SARS-CoV-2-positive study group patients, there were seven symptomatic patients and fourteen asymptomatic patients. There were no significant differences between symptomatic and asymptomatic patients in the number placentas with SGA (5/7 vs. 9/14), MVM(4/7 vs 6/14), FVM (4/7 vs 6/14), or signs of chronic inflammation reflected by chronic deciduitis or VUE (2/7 vs 3/14).

## Discussion

*Placental pathology with SARS-CoV-2 positive women*

The data show that we must accept the null hypothesis that there is no significant difference in placental pathology in women tested for SARS-CoV-2, comparing those positive with those negative. This differs from two case-control studies of placental pathology in patients who were *SARS-CoV-2 positive*. Shanes and associates examined15 placentas delivered in the third trimester in COVID positive women and reported a high prevalence of decidual arteriopathy and other features of MVM.[10] This study included a 16-week intrauterine fetal demise in which fetal tissue tested negative for SARS-CoV-2.[10] Smithgall and colleagues demonstrated that 52 third trimester placentas from SARS-CoV-2-positive women had higher prevalence of villous agglutination and subchorionic thrombi, compared to placentas in the control group.[11] None of these pathologies were more frequent in our two groups of patients were compared based on COVID-19 test-positivity. In a case series, Baergen and Heller found that ten of 20 (10/20, 50%) cases of COVID positive patients showed some evidence of FVM or fetal vascular thrombosis (FTV).[12] Moreover, Hecht and colleagues examined placentas of nineteen COVID-19 exposed women and found that there was no characteristic histopathology present among the 19 cases, despite two placental infections.[13]

Our study included a control group of placentas to compare with placentas from SARS-CoV-2 positive women, and we did not identify a specific pathology or pathological pattern attributed to the virus. Compared to placentas from SARS-CoV-2-negative women, there were no significant differences in gross and microscopic placental pathology, including placental weight, SGA placentas, abnormal cord insertion or cord coiling, MVM, FVM, and maternal or fetal inflammatory responses. Our results of no signature pathology for placentas in women positive for COVID-19 reflect the findings of our local obstetrical patient population in St. Louis, MO, from which the study subjects were derived. Populations from other locales, such as the ones reported in the above papers, may exhibit different placental responses that reflect regional differences in patient populations.

The microscopic findings from current study support our acceptance of the null hypothesis that SARS-CoV-2 infection acquired during third trimester does not have a demonstrable impact on placental structure and pathology. Notably, our acceptance of the null hypothesis is further supported by absence of significant differences in placental weight and fetal/placental weight ratio. Placental weight is a surrogate for placental growth and development while fetal/placental weight ratio is often regarded as a surrogate marker for placental efficiency.

We do not rule out a greater impact if the infection occurred earlier in pregnancy.

***Key issues related to pregnancy associated with COVID-19***

*ACE2 in placenta*

Immunocytochemical expression of ACE2 is localized in the syncytiotrophoblast, cytotrophoblast, endothelium and vascular smooth muscle of primary and secondary villi.[6]

Expression of ACE2 by immunohistochemistry in placentas of COVID-19 exposed pregnancies showed ACE2 membranous expression polarized in the villous syncytiotrophoblast (with immunoreactivity highest on the stromal side of the ST). Importantly, there was also cytotrophoblast and extravillous trophoblast expression of ACE2. In contrast, no ACE2 expression was detected in villous stroma, Hofbauer cells, or endothelial cells.[13] ACE2 protein is enriched per unit volume in the placenta and ovary, compared to other tissues, despite a low mRNA level detected in the two tissues.[14] The studies offer a potential mechanism for maternal-fetal transmission.

*Does SARS-CoV-2 infect placenta?*

Smithgall and colleagues found no evidence of direct viral infection of the placenta, using in-situ hybridization for the S-gene encoding the spike protein and immunohistochemistry with the monoclonal-SARS-CoV-2 spike-antibody 1A9.[11] Conversely, in two of 19 cases reported by Hecht et al.[13], SARS-CoV-2 RNA was present in placental villi focally, in both the ST and cytotrophoblast.[13] Moreover, three of 11 placentas having swabs taken from villi and fetal membranes tested positive for SARS-CoV-2.[15] An additional case report, of a woman who was infected with SARS-CoV-2 at 22 weeks gestation, showed that the placenta and umbilical cord were positive for SARS-CoV-2 at delivery.[16] Localization of SARS-CoV-2 spike protein and RNA was found in both villi and peri-villous fibrin.[16] Collectively, these data suggest that viral colonization of the placenta and fetal membranes may occur but not in a predictable manner. Whether colonization leads to infection remains unknown.

*Is there vertical transmission*

Studies of vertical transmission to date have shown conflicting results,[2,17] but a cautious conclusion is that there is minimal, if any, role for vertical transmission of the virus through the placenta to the fetus. Clearly, this is a key question in understanding the impact of COVID-19 on maternal and neonatal morbidity and mortality.

*Placental pathology in SARS and MERS*

Verma and colleagues reviewed placental pathology associated with SARS or MERS infection.[2] Placentas from these coronavirus viruses show acute and chronic placental insufficiency and associated with intrauterine growth restriction (IUGR) or miscarriage in 40% of affected pregnancies.[18-20] However, the key issue again lies in the need for studies with a control group. A case in point is our study, which demonstrated almost half of the cases in the study group demonstrated features of maternal or fetal vascular malperfusion; however, there was no statistically significant difference in the study group when compared with the control group.

**Conclusion**

Third trimester placentas from SARS-CoV-2-positive women do not express specific pathology or pathological pattern. While they harbor a relatively high prevalence of features of maternal or fetal vascular malperfusion, these features were not unique as they were also present in women who had a clinical indication for SARS-CoV-19 testing with test-negative results. Pregnancy complicated with COVID-19 during the 3$^{rd}$ trimester does not have an obvious effect on placental structure and pathology. We do not rule out a greater impact if the infection occurred earlier in pregnancy.

**Funding:**

This study was supported by the departmental faculty development to Dr. Mai He from the Department of Pathology & Immunology, Washington University in St. Louis School of Medicine, St. Louis, MO 63110, USA

**Disclosure:**

All authors disclose no conflicts of interest. IUM serves on the Scientific Advisory Board of Luca Biologics.

**Table 1. Clinical and pathological information of study and control groups**

| | Study group (SARS-CoV-2 tested positive) N = 21 | Control group (SARS-CoV-2 tested positive) N = 20 | P value |
|---|---|---|---|
| Maternal age (years, minimal to maximum, median) | 17-37, 30 | 18-37, 27 | NS |
| Gestational age at delivery (weeks, minimal to maximum, median) | 33-40, 37.5 | 28-41, 38 | NS |
| Placental weight (grams, minimal to maximum, median) | 337.5-577, 446 | 255-632 432 | NS |
| Fetal/placental weight ratio (mean, std dev) | 6.64 ±0.83 | 6.58 ±1.53 | NS |
| Features of maternal vascular malperfusion | 10 (47.6%) | 4 (20%) | NS |
| Features of fetal vascular malperfusion | 10 (47.6%) | 8 (40%) | NS |
| Maternal inflammatory response | 2 (9.5%) | 3 (15%) | NS |
| Fetal inflammatory response | 2 (9.5%) | 4 (20%) | NS |
| Chronic inflammation (chronic deciduitis and VUE) | 5 (23.8%) | 5 (25%) | NS |

NS: No significant difference.
VUE: Villitis of unknown etiology

**Table 2. Placental pathology in placentas from SARS-CoV-2-positive women (the study group)**

| Patient No. | 1 | 2 | 3 | 4 | 5 | 6 | 7 | 8 | 9 | 10 | 11 | 12 | 13 | 14 | 15 | 16 | 17 | 18 | 19 | 20 | 21 | N (%) | P |
|---|---|---|---|---|---|---|---|---|---|---|---|---|---|---|---|---|---|---|---|---|---|---|---|
| Age (y) | 35 | 27 | 27 | 27 | 18 | 32 | 30 | 31 | 20 | 17 | 21 | 35 | 31 | 19 | 34 | 19 | 30 | 31 | 37 | 31 | 34 | | NS |
| Gestational age at delivery (W) | 39 | 39 | 39 | 37 | 39 | 37 | 39 | 40 | 39 | 37 | 37 | 33 | 38 | 34 | 37 | 36 | 39 | 39 | 37 | 37 | 37 | | NS |
| Respiratory symptoms | | X | | | | | | | X | X | | X | | | X | | | X | | X | | 7 (33.3) | |
| Maternal conditions during this pregnancy | Anemia, asthma | fever and cough | "GBS+, PN, flu | N | N | CHTN, sarcoidosis | GDM | GBS+ | HT, fever | Loss of taste and smell | PE | Pneumonia, RF | GHTN, IDA | CHTN | RA, pneumonia | HSV | GHTN | Asthma, GBS+, fever | PE | Tachy-pnea | HCV, HSV, Substance abuse | | |
| SGA placenta | X | X | | X | X | X | X | | X | X | X | | | X | X | X | | | X | X | | 14 (66.7) | NS |
| LGA placenta | | | | | | | | | | | | | | | | | | | | | | | N/A |
| **Features of Maternal vascular malperfusion** | | X | | | | X | | | | X | X | | | X | | X | X | X | X | X | | 10 (47.6) | NS |
| Decidual vasculopathy | | | | | | | | | | X | X | | | X | | | | | | X | | | NS |
| Retroplacental hematoma | | | | | | X | | | | | | | | | | | X | | | | | | NS |

| | | | | | | | | | | | | | | | |
|---|---|---|---|---|---|---|---|---|---|---|---|---|---|---|---|
| Trophoblastic giant cells | | X | | | X | X | X | X | | X | | | | | NS |
| Parenchymal infarct | | | | | | | | | | | X | | | | NS |
| Accelerated villous maturity | | | | | | | X | X | | | | | | | NS |
| Villous agglutination | | | | | | | | | | | | | | | NS |
| Abnormal cord insertion | X | | | | | | X | | | | | X | | | NS |
| Hypercoiling of umbilical cord | | | | | | | | | | | | | | | NS |
| True knot of cord | 1 | | | | | | 14 | | | | | | 21 | | NS |
| **Features of fetal vascular malperfusion** | | X | X | X | X | | X | X | X | X | | X | X | 10 (47.6) | NS |
| Avascular villi, clustered | | | | | | | X | | | | | | X | | NS |
| Thrombi of fetal vessels | | | | | | | | | | | | | | | NS |
| Hemorrhagic endovasculitis | | | | | | | | | | | | | | | NS |
| Ectasia of chorionic plate fetal vessels | | X | X | | | | X | | | X | | | | | NS |
| Chorangiosis | | | | X | X | | X | X | X | | | X | | | NS |

| | | | | | | | | | | | |
|---|---|---|---|---|---|---|---|---|---|---|---|
| Delayed villous maturation | | | | | X | | | | | | |
| Villous edema | | | | | | X | X | | | | NS |
| **Maternal inflammatory response** | | X | | X | | | | | | 2 (9.5) | NS |
| **Fetal inflammatory response** | | X | | X | | | | | | 2 (9.5) | NS |
| Acute chorioamnionitis | | X | | X | | | | | | | NS |
| Acute vasculitis or funisitis | | X | | X | | | | | | | NS |
| Chronic inflammation | X | | X | | | | X | X | X | 5 (23.8) | |
| Chronic deciduitis with plasma cells | X | | X | | | | X | X | | | NS |
| VUE | | | | | | | | | X | | NS |
| **Other placental pathology** | | | | | | | | | | | |
| Intervillous thrombi | | | | | | | | | | | NS |
| Increased perivillous | | | | | X | | | | | | NS |

| | |
|---|---|
| fibrin/fibrinoid deposition | |

Note: N: Normal or none. NS: No significant difference.
CHTN: Chronic hypertension. GBS: Group B streptococcus. GDM: Gestational diabetes mellitus. GHTN: Gestational hypertension. HT: Hypothyroidism. IDA: Iron-deficient anemia. PE: Preeclampsia. PN: Pyelonephritis. RF: Respiratory failure.